\documentclass[reprint,prx,amsmath,amssymb,aps,notitlepage]{revtex4-2}

\usepackage[linesnumbered, ruled]{algorithm2e}
\usepackage{amsmath} 
\usepackage{algorithmic} 
\usepackage{graphicx}
\usepackage{url}
\usepackage{hyperref}
\usepackage[hyphenbreaks]{breakurl}

\usepackage{fancyhdr}

\def\half{ {\frac{1}{2}} }

\renewcommand{\v}[1]{\boldsymbol{#1}}

\begin{document}

\title{Continuous body 3-D reconstruction of limbless animals}

\author{Qiyuan Fu$^1\dagger$}
\author{Thomas W. Mitchel$^1\dagger$}
\author{Jin Seob Kim$^1$}
\author{Gregory S. Chirikjian$^{2,1}$}
\author{Chen Li$^{1,\ddagger}$}
  \homepage{https://li.me.jhu.edu/}

\affiliation{$^1$Department of Mechanical Engineering, Johns Hopkins University, Baltimore, MD 21218, USA}
\affiliation{$^2$Department of Mechanical Engineering, National University of Singapore, 117575, Singapore}

\collaboration{$^{\dagger}$Equal contributions; $^\ddagger$Corresponding author: chen.li@jhu.edu}

\date{First posted online 3 February 2021 as 10.1242/jeb.220731}

\begin{abstract}
 Limbless animals such as snakes, limbless lizards, worms, eels, and lampreys move their slender, long body in three dimensions to traverse diverse environments. Accurately quantifying their continuous body's 3-D shape and motion is important for understanding body-environment interaction in complex terrain, but this is difficult to achieve (especially for local orientation and rotation). Here, we describe an interpolation method to quantify continuous body 3-D position and orientation. We simplify the body as an elastic rod and apply a backbone optimization method to interpolate continuous body shape between end constraints imposed by tracked markers. Despite over-simplifying the biomechanics, our method achieves a higher interpolation accuracy (\string~50\% error) in both 3-D position and orientation compared with the widely-used cubic B-spline interpolation method. Beyond snakes traversing large obstacles demonstrated, our method applies to other long, slender, limbless animals and continuum robots. We provide codes and demo files for easy application of our method.
\end{abstract}


\maketitle

\section{Introduction}

Limbless animals such as snakes \citep{Byrnes2012,Gans1986,Goldman2010,Gray1950kinetics,Jayne1985,Jayne2013a,Lillywhite2000,Marvi2014,Munk2008,Socha2002}), limbless lizards \citep{Gasc1990, Gans1992, Miller1944}, worms \citep{Karbowski2006, Park2008, Summers1997, Dorgan2015a}, and fish \citep{Gillis1998, Jayne1995, Tytell2007, Herrel2011, Gidmark2011, Gemmell2015} can deform their long, slender body to move through a large diversity of environments. To move through complex environments such as branches \citep{Jorgensen2017, Byrnes2012}, underwater sand beds \citep{Gidmark2011,Tatom-Naecker2018}, large obstacles \citep{Gart2019}, and even during gliding \citep{Socha2011a} and swimming \citep{Graham1987}, their body must deform substantially ($>$ 10\% body length) in three dimensions. In addition, how the body is oriented and rotates locally relative to the environment often strongly affects the forces that the animal generates. For example, snakes, limbless lizards, and worms all have bodies with anisotropic friction properties \citep{Hu2009,Spinner2015,Schulke2011,Shen2012}. During borrowing in granular media, granular resistive stress depends sensitively on local orientation of a moving body \citep{Li2013}. In addition, pressure drag and skin friction in fluids depends on body shape and orientation \citep{Tytell2007,Gemmell2015,Taylor1952,Jayne1995,Holden2014,Danos2012}. Furthermore, when traversing large obstacles, where the body contacts the terrain affects a limbless animal's stability \citep{Gart2019}. Therefore, to understand limbless locomotion in these environments, it is important to quantify an animal's body shape and movement both in position and orientation in three dimensions accurately.

Many previous studies quantify limbless animal body shape and movement in three dimensions by a series of tracked points \citep{Kwon2013,Byrnes2012,Marvi2014,Socha2005}. However, discrete points cannot accurately capture curved local body shape necessary for quantifying body-environment interaction (e.g., Movie 1) unless a very large number of markers are tracked, which is time-consuming and challenging when marker occlusion is frequent (e.g., large 3-D rotation, multiple obstacles present). Some studies use superposition of curves generated by basis functions (e.g., B-spline) \citep{Sharpe2015,Schiebel2018,Fontaine2008,Padmanabhan2012,Yeaton2020}, but the majority of these superposition methods were for planar locomotion. In addition, none of these geometric interpolation methods above capture body rotation (roll) about the longitudinal axis (although some studies measured it using computer vision techniques \citep{Donatelli2017} or manually \citep{Fish2007}).

A simplistic way of quantifying a long, slender body, with both position and orientation information, is to approximate it as an elastic rod. Elastic rod theories developed by Kirchhoff and Cosserat view an elastic rod as continuum on which forces are applied (for reviews, see \cite{Cao2008a,Dill1992b,Zhang2019}). Despite being biomechanically over-simplified---not considering musculoskeletal morphology and muscle function \citep{Dickinson2000} (but see \cite{Zhang2019})---elastic rod modeling has facilitated first-order modeling and basic understanding of the mechanics (e.g., muscle activation, resistive forces from the surrounding media) of limbless animals moving both in two \citep{JohnH.Long1998,Tytell2018,ding2013emergence,Cheng1998,Zhang2019} and three \citep{McMillen2006} dimensions.

\begin{figure*}[!t]
  \centering
  \includegraphics[width=\textwidth]{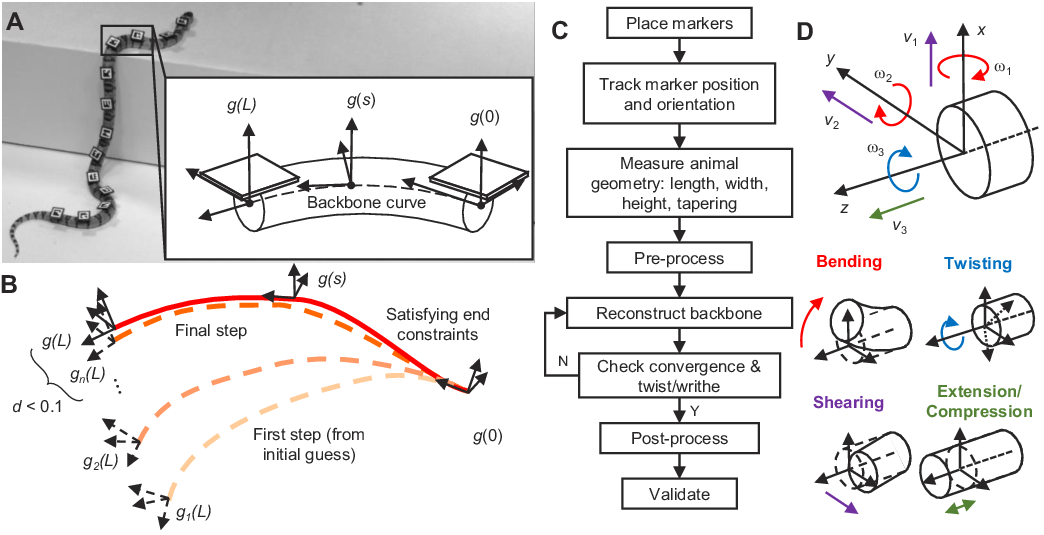}
  \caption{\textbf{Overview of interpolation method combining elastic rod modeling and backbone optimization.} (A) Modeling of a snake body segment as an elastic rod segment, with its midline described by a backbone curve $g(s)$ (dashed). The rod segment is subject to two end constraints imposed by 3-D position and orientation of two body frames, $g(0)$ and $g(L)$, obtained from markers attached to it at both ends. $L$ is segment length. $g(s)$ is a body frame at length $s$. (B) Schematic of inverse kinematics to converge backbone curve towards marker end constraints. See \textbf{Inverse Kinematics} for description of the process. (C) Workflow to use method. (D) Definition of different types of elastic rod deformation: lateral and dorsoventral bending about $x$ and $y$ axes, twisting about $z$ axis, lateral and dorsoventral shearing along $x$ and $y$ axis, and extension or compression along $z$ axis. $x$-$y$-$z$ axes form a right-handed body frame attached to backbone curve, with +$x$ axis pointing upward along the axis of left-right symmetry of cross section, and +$z$ axis pointing backward tangent to segment backbone curve.}
  \label{fig_overview}
\end{figure*}

In the field of robot motion planning, elastic rod modeling has been combined with backbone optimization to generate motion trajectories of snake-like, high degree-of-freedom (hyper-redundant) robots to achieve the desired locomotor tasks \citep{Burdick1994,Chirikjian2015}. The idea of backbone optimization is that, when the position and orientation of both ends of a high degree-of-freedom system are given (hereafter referred to as end constraints), the approximate shape (referred to as the backbone curve, Fig. \ref{fig_overview}A inset, dashed curve) of its actual midline in between (referred to as the midline) can be determined by minimizing a cost function. By representing a long, slender body as an elastic rod with its elastic potential energy as the cost function and applying backbone optimization over time, one can determine the evolution of body shape, or motion trajectory, that satisfies the desired end constraints (e.g., moving from point A to point B around an obstacle). Similarly, this method has enabled shape interpolation of stiff molecules such as DNA subject to end constraints \citep{Kim2006a}. It is important to note that, in backbone optimization, the use of an elastic rod is not intended to capture the physical reality of the system, but instead to simplify mathematical calculations (see \textbf{Overview}).

Here, we propose to use this method combining elastic rod modeling and backbone optimization to reconstruct the continuous body of limbless animals in three dimensions, using end constraints from discrete tracked markers. We emphasize that the purpose of our method here is to geometrically quantify the animal's continuous body shape and movement kinematics, rather than model its locomotion mechanics and dynamics (for doing this instead, see \cite{Boyer2010}, \cite{Zhang2019}). We describe our method using the example of locomotion of the variable kingsnake (\textit{Lampropeltis mexicana thayeri} \citep{garman1883reptiles}) with large body deformation and motion in three dimensions. We simplify the animal body as an elastic rod and apply backbone optimization to interpolate between markers placed on the body to obtain 3-D position and orientation of the continuous body. In addition, we characterized how marker separation affects positional accuracy of the interpolation, using a ground truth of the body midline positions extracted via computer vision techniques. Further, we compared our method with the widely-used B-spline method (e.g., \cite{Sharpe2015,Schiebel2018,Fontaine2008,Yeaton2020}) by applying both to a large dataset of kingsnakes traversing large step obstacles and comparing their position and orientation errors using tracked marker data as reference. We first interpolated the $x, y, z$ positions together by fitting a 3-D B-spline curve to tracked markers locations, then calculated orientation from the interpolated B-spline position curve using the Bloomenthal method \citep{Bloomenthal1990}. Finally, we discuss why our method achieves higher interpolation accuracy than entirely geometric interpolation methods and the benefits of such higher accuracy. We provide MATLAB codes and demonstration files for users to easily learn and apply our method to their studies (see
\textbf{Data availability}).

\section{Materials and Methods}

\subsection{Overview}

The model system to describe our method is the variable kingsnake traversing large step obstacles \citep{Gart2019}. Multiple markers are attached to the animal body from head to tail to capture both 3-D position and orientation locally, and they divide the body into segments (Fig. \ref{fig_overview}A). Each elastic rod segment approximating a body segment is subject to two end constraints imposed by the 3-D position and orientation of the two markers at both its ends (Fig. \ref{fig_overview}A, insets). Interpolation is performed piecewise between each adjacent pair of markers and for both 3-D position and orientation, which is simplified using Lie group theory (with the number of equations reduced and singularity of Euler angles avoided). For an introduction of Lie group theory, see \cite{murray2017mathematical}. Below we follow the convention in \cite{Kim2006a}.

Our method approximates the continuous shape of each body segment by that of a quasi-static elastic rod dominated by elastic forces, subject to end constraints from tracked markers. This is a drastic over-simplification, because in reality the deformation of an animal body segment is a result of all the forces acting on it, which include gravitational force, inertial force due to body acceleration, forces from the surrounding environment, and internal forces due to muscle activation and viscoelastic skin, tissue, and skeleton \citep{Cheng1998}. The purpose of this over-simplification is to simplify derivation of the Lagrangian (such that it is merely the elastic potential energy of the rod), which we used in backbone optimization to interpolate between end constraints. In principle, other forces can be modeled by the Lagrangian (e.g., gravitational potential energy) or as constraints using Lagrangian multipliers (e.g., non-conservative forces). However, as we demonstrate in \textbf{Results}, even with such a drastic over-simplification, our method achieves higher accuracy in 3-D shape interpolation than the widely-used, purely geometric B-spline method.

In our method, the elastic rod can experience twisting, shear, extension, and compression aside from bending. It is unclear whether these body deformation types occur in every species---for example, twisting exists in some snakes but is limited by the vertebrae structure and varies between species \citep{Moon1999, Jurestovsky2020}. However, inclusion of these deformation types in the model allows more general application to different species. It also allows for better interpolation by accommodating the inevitable measurement errors of end constraints and segment lengths due to camera tracking noise and movement of the animal's skin. The reconstruction results with lateral and dorsoventral bending having the dominant contributions to elastic energy (Fig. S2) suggested that the inclusion is reasonable when applied to snakes.

For a long, slender body with very high degrees of freedom to satisfy a few discrete end constraints, many solutions exist. This is referred to as the kinematically redundant or hyper-redundant problem \citep{Chirikjian1995}. A well-developed method to address this problem is backbone optimization, which minimizes a cost function to determine one or a few solutions \citep{Chirikjian1995}. Such constrained optimization problems are usually solved using the Euler-Lagrange equation, which requires the use of coordinates that can result in singularity. Here, we use a coordinate-free method to avoid singularities, which consists of the Euler-Poincar\'{e} equation and a kinematic equation \citep{Chirikjian2015}, with the elastic energy of the rod segment as the cost function to minimize, subject to end constraints. This results in a solution of spatial velocity (spatial derivative of body frame, as viewed in the instantaneous local body frame; see Eqn. (\ref{eqn_xi})), which can be integrated from one end constraint towards another to obtain a backbone curve.

In order for the integrated backbone curve to interpolate between the two end constraints measured by adjacent markers, we apply the method of inverse kinematics \citep{Kim2006a} (Fig. \ref{fig_overview}B). For each body segment, we first start from one end attached to one of the measured end constraints and integrate spatial acceleration (spatial derivative of spatial velocity) to obtain a backbone curve. The spatial acceleration is calculated using an initial guess of spatial velocity at the starting end and the Euler-Poincar\'{e} equation. We then iteratively perturb the backbone curve (by perturbing the initial guess of spatial velocity) to reduce the distance of its other end to the second measured end constraint until they converge. This entire procedure is done piecewise for all segments between adjacent markers, and the interpolated backbone curves of all segments together give the approximate continuous midline of the entire body (except the very front and rear end without markers).

To apply our method in animal experiments, the workflow is as follows (Fig. \ref{fig_overview}C). First, users need to place on the animal body markers that can provide 3-D position and orientation information, such as BEEtags \citep{Crall2015}, ArUco \citep{Garrido-Jurado2014}, and custom rigid body markers \citep{Ravi2013}). The users need to then track the markers to obtain their \textit{x-y-z} coordinates and Euler angles as input. Then, they need to measure animal length, width, and height (and body tapering if it is substantial) and pre-process the tracking data to obtain end constraints. After these preparations, they can run the backbone interpolation codes followed by post-processing to obtain the interpolated animal body midline for further analysis. Convergence check and twist/writhe (how much the rod twists into coils) check are performed automatically before results are saved. To validate the results, the users can project the reconstructed body midline onto videos, visually check the match, and fine-tune interpolation parameters to improve accuracy.

Below we describe our method in detail. Because our interpolation is done piecewise on the body, the description focuses on one body segment between two adjacent markers approximated by one elastic rod segment, unless specified otherwise.

\subsection{Backbone curve}

The backbone curve describes the spatial and temporal evolution of a series of body frames along the body segment (Fig. \ref{fig_overview}A), placed sufficiently closely to each other to obtain a (near) continuous description of the body shape and motion in three dimensions. Each body frame has its spatial configuration---translation and rotation relative to a fixed world frame---represented by a matrix:
\begin{equation} \label{body_frame}
  g(s, t) =
\begin{bmatrix}
  R(s,t) & {\v p}(s,t) \\
  {\v 0}^T & 1
\end{bmatrix}
\end{equation}
where $R(s,t) = \begin{bmatrix} {\v r}_1(s,t) & {\v r}_2(s,t) & {\v r}_3(s,t) \end{bmatrix} $ is a $3\times3$ matrix representing the marker orientation, with each column the coordinates of a unit vector along the +$x, y, z$ axis of the body frame in the fixed world frame (Fig. \ref{fig_overview}D), ${\v p}(s,t) = \begin{bmatrix} x(s,t) & y(s,t) & z(s,t)\end{bmatrix}^T$ are the $x, y, z$ coordinates of the origin of the body frame, ${\v 0}^T = \begin{bmatrix}0 & 0 & 0\end{bmatrix}$, ${M}^T$ denotes the transpose of a matrix $M$, $s$ is arc length from one end of the unstretched body segment (e.g., the marker closest to snout or tail tip), and $t$ is time. Below, all lengths refer to unstretched lengths unless stated otherwise.

\subsection{Elastic energy of rod}

An elastic rod can be bent, twisted, sheared, extended, or compressed under external forces and torques (Fig. \ref{fig_overview}D). Below we derive the elastic energy of a rod segment using Lie group notations, which will be used for optimization in the next section \citep{Kim2006a}.

The spatial velocity (unit-length spatial deformation) of the backbone curve is described by a vector:
\begin{equation} \label{eqn_xi}
{\v \xi}(s, t) = \begin{pmatrix} {\v \omega}(s,t) \\ {\v v}(s,t) \end{pmatrix} = \left(g(s,t)^{-1}\frac{\partial g(s,t)}{\partial s}\right)^{\vee}
\end{equation}
in which ${\v \omega} = \left[\omega_1, \omega_2, \omega_3\right]^T$ is the unit length rotational deformation (bending and twisting), ${\v v} = \left[v_1,v_2,v_3\right]^T$ is the unit length translational deformation (shearing and extension/compression), $g^{-1}$ denotes the inverse of matrix $g$, and ${}^{\vee}$ is an operator that extracts a 6-dimensional vector from a $4\times4$ matrix:
\begin{equation}
{\begin{bmatrix}0 & -\omega_3 & \omega_2 & v_1 \\ \omega_3 & 0 & -\omega_1 & v_2 \\ -\omega_2 & \omega_1 & 0 & v_3 \\ 0 & 0 & 0 & 0\end{bmatrix}}^{\vee} = \begin{bmatrix} \omega_1 \\ \omega_2 \\ \omega_3 \\ v_1 \\ v_2 \\ v_3\end{bmatrix}
\end{equation}

Then, the evolution of elastic forces and torques is given by:
\begin{eqnarray} \label{muscle_wrench}
  {\v F}(s,t) = K ({\v \xi}(s,t)-{\v \xi}_0)
\end{eqnarray}
where $K$ is the $6\times6$ stiffness matrix, which is symmetric ($K = K^T$) and assumed to be constant, ${\v \xi}_0$ is the intrinsic deformation of the body frame along the rod when there is no external force. We further assumed that there is no coupling between forces along different directions and torques about different axes, i.e., $K$ is diagonal.

Then, the elastic energy within an elastic rod segment of length $L$ at a given time $t$ is:
\begin{eqnarray} \label{work_body}
  \begin{aligned}
  E(t) & = \int_0^L{\half{({\v \xi}(s,t)-{\v \xi}_0)^{T}K({\v \xi}(s,t)-{\v \xi}_0)}} \ ds \\
  & = \int_0^L{(\half{{{\v \xi}(s,t)}^{T}K{\v \xi}(s,t)}-{\v k}^T{\v \xi}(s,t)+\beta'}) \ ds \\
  \end{aligned}
\end{eqnarray}
where ${\v k} = K{\v \xi}_0$ and $\beta' = \half{{\v \xi}_0^TK{\v \xi}_0}$ are two constants.

\subsection{Backbone optimization}

For an elastic rod segment that satisfies any two end constraints, we can obtain its backbone curve using backbone optimization with elastic energy as the cost function. We omit time $t$ in this and next sections considering the quasi-static assumption, which works reasonably for friction-dominated snake locomotion \citep{Hu2009}. For more details of the backbone optimization method, see \citep{Kim2006a}.

Using the Euler-Poincar\'{e} equation widely used to solve constrained optimization problems \citep{Chirikjian2015}, we derive a kinematic equation that describes the optimal solution, which is an ordinary differential equation:

\begin{equation} \label{xi_diff_eqn}
  K\frac{d{\v \xi}}{ds} + (K{\v \xi}-{\v k})\wedge{\v \xi}=0
\end{equation}
where $\wedge$ is an operator defined by:
\begin{equation}
 \begin{pmatrix}{\v \omega}_1 \\ {\v v}_1 \end{pmatrix} \wedge \begin{pmatrix}{\v \omega}_2 \\ {\v v}_2 \end{pmatrix} = \begin{pmatrix}{\v \omega}_2\times{\v \omega}_1 + {\v v}_2\times{\v v}_1 \\ {\v \omega}_2\times{\v v}_1 \end{pmatrix}
\end{equation}

Next, we numerically integrate Eqn. (\ref{xi_diff_eqn}) to obtain spatial velocity ${\v \xi}(s)$. Then, starting from one end of the segment ($s = 0$) with a spatial velocity value chosen at this end ${\v \xi}(0)$ (boundary condition), one can integrate ${\v \xi}(s)$ over arc length $s$ towards the other end ($s = L$) to obtain a backbone curve $g(s)$ of the entire segment ($s\in\left[0,L\right]$, where $L$ is segment length). The resulting backbone curve depends on the boundary condition ${\v \xi}(0)$ chosen and does not necessarily satisfy the measured end constraints from tracked markers. Next, we apply the method of inverse kinematics to converge the backbone curve to satisfy the measured end constraints.

\subsection{Inverse kinematics}

In the first iteration ($k = 1$), we start from one end constraint $g(0)$ tracked by a marker (Fig. \ref{fig_overview}B) and perform the integration above to obtain a backbone curve that ends at $g_1(L)$. Because the markers do not provide information about spatial velocity, in the first step, we use an initial guess of $\v \xi(0) = \v \xi_1(0)$. As a result, the other end $g_1(L)$ of the resulting backbone curve (Fig. \ref{fig_overview}B, bottom dashed curve) is far away from the other end constraint from the other tracked marker $g(L)$.

Over the following steps $k = 2, ..., n$, we start with ${\v \xi}_{k-1}(0)$ used in the previous step and iteratively perturb it to perturb the backbone curve. Using the inverse kinematics method, we gradually reduce the distance from the other end $g_k(L)$ of the resulting backbone curve to the other measured end constraint $g(L)$, until they converge. This results in the final backbone curve (Fig. \ref{fig_overview}B, top dashed curve) that converges to the backbone curve that exactly satisfies the two end constraints from tracked markers (Fig. \ref{fig_overview}B, solid curve).

Because our goal is to minimize interpolation error in both 3-D position and orientation, we use the weighted Frank Park distance \citep{Park1995} between $g_k(L)$ and $g(L)$ to provide a measure of how close two body frames are (position and orientation combined) in Lie group space, defined as:
\begin{equation} \label{Frank Park distance}
  d(g_k(L),g(L)) = \|{R_k(L)-R(L)}\|_F+\|{\v p}_k(L)-{\v p}(L)\|
\end{equation}
where $\|{R_k(L)-R(L)}\|_F$ is the Frobenius norm of a matrix $R_k(L)-R(L)$ that measures error in orientation, and $\|{\v p}_k(L)-{\v p}(L)\|$ is the Euclidean norm of a vector ${\v p}_k(L)-{\v p}(L)$ that measures error in position.

We set the inverse kinematics iteration to stop when $d(g_n(L),g(L))<0.1$. This results in a position error smaller than 0.1 mm (1\% body diameter of 9 mm) and an orientation error smaller than 6$^\circ$. If not, the iteration stops when the maximal iteration step of 65 is reached. For more details of the inverse kinematics iteration, see \citep{Kim2006a}.

Sometimes, an initial guess of $\v \xi(0)$ does not produce a backbone curve that converges to the tracked marker end constraint $g(L)$ or results in a shape with unrealistic, large twist or writhe. To resolve this, we try different initial guesses. We try the inverse kinematics procedure at most three times for each trial before a solution is found for each body segment in each video frame, using different initial guesses, iteration step size, and maximal number of iterations. We also perform a twist and writhe check \citep{Kim2006a} using pre-defined thresholds of 0.4 (determined by visually examining body frames interpolated using different thresholds projected onto the videos) to ensure that the interpolated backbone curve is realistic.

To reduce computation time, we always first try the ${\v \xi}(0)$ value from the previous successful video frame for the same segment for the following frame, because the shape change between these video frames is usually small with a sufficiently high camera frame rate. If no previous video frame succeeds or this ${\v \xi}(0)$ value fails to produce realistic curves within the maximal iteration steps twice, we then try pre-defined initial guesses with extension and bending in different directions.

\subsection{Interpolation parameters}

\subsubsection{Segment lengths, cross-sectional shape, and tapering}

Several geometric parameters in the method affect the fidelity of the interpolation and should be measured as input, including the length of the body segment between markers $L$, body width and height, and body tapering.

Ideally, segment lengths should be directly measured piecewise between each pair of adjacent markers after marker placement. Because this was challenging for the conscious animals in our experiments, we approximated segment lengths by the largest distance observed between adjacent markers across all video frames in all trials with the same marker placement. Digital photography analysis is an alternative method \citep{Astley2017}. During reconstruction, we further fine-tuned segment lengths to improve visual match by projecting the reconstructed body onto experimental videos and evaluating whether the estimated segment lengths were too long or too short and adjusting them accordingly.

We measured the width and height of the three kingsnakes at 10 nearly equally spaced locations along the body (Fig. S1). We found that their body width is similar to body height, i.e., the body cross section is nearly circular. In addition, there is little body tapering. Based on these measurements, for interpolation, we approximated the cross-sectional shape as a circle with a radius $r = 4.5$ mm equal to the average of width and height measurements. For pre-processing of marker data, we calculated the body radius of each individual as a function of body length, $r(s)$, by linear interpolation.

\subsubsection{Stiffness matrix}

As mentioned in Eqn. (\ref{muscle_wrench}), the stiffness matrix $K$ is diagonal:
\begin{equation}
K = \begin{bmatrix}
  \kappa_1 & 0 & 0 & 0 & 0 & 0 \\
  0 & \kappa_2 & 0 & 0 & 0 & 0 \\
  0 & 0 & \kappa_3 & 0 & 0 & 0 \\
  0 & 0 & 0 & \kappa_4 & 0 & 0 \\
  0 & 0 & 0 & 0 & \kappa_5 & 0 \\
  0 & 0 & 0 & 0 & 0 & \kappa_6
\end{bmatrix},
\end{equation}
where $\kappa_1$ and $\kappa_2$ are lateral and dorsoventral bending stiffnesses about $x$ and $y$ axes, $\kappa_3$ is torsional stiffness, $\kappa_4$ and $\kappa_5$ are lateral and dorsoventral shearing stiffnesses along the $x$ and $y$ axes, and $\kappa_6$ is extensive/compressive stiffness (which are assumed to be the same).

Many limbless animals have body cross-sectional shapes that are elliptical or irregular (e.g., \cite{Donatelli2017}, \cite{Holden2014}). Considering that the kingsnake's body cross section is nearly circular, we further assumed that the elastic rod has the same stiffness parameters along the $x$ and $y$ axes considering the symmetry of a circular cross section:
\begin{align}
  \begin{split}
    \kappa_1 = \kappa_2 &= E I \\
    \kappa_3 &= G J\\
    \kappa_4 = \kappa_5 &= G A \\
    \kappa_6 &= E A
  \end{split}
\end{align}
where $I$ and $J$ are second area moments of inertia about $x$ (or $y$) axis and $z$ axis (for asymmetric cross section $I$ is different around $x$ from that around $y$ axis), $E$ and $G$ are Young's and shear moduli, and $A$ is cross-sectional area of the rod. Shear modulus can be calculated from Young's modulus via:
\begin{equation}
  G = \frac{E}{2(1+\nu)}
\end{equation}
where $\nu$ is the Poisson's ratio.

With these assumptions, all elements of $K$ scale with $E$. As a result, changing $E$ simply scales elastic energy and does not change the solution.

With the circular cross section approximation, we calculated the second moments of inertia $I$ and $J$ and cross-sectional area $A$ as:
\begin{align}
  \begin{split}
    I &= \frac{\pi}{4}r^4 \\
    J &= \frac{\pi}{2}r^4 \\
    A &= \pi r^2
  \end{split}
\end{align}
If the cross section is not circular, such as an ellipse, these moments of inertia need to be calculated accordingly. Our preliminary tests showed that the interpolated midline only changed by 6\% body diameter in position when width:height ratio changed from 1 to 0.3, although buckling appeared when it reduced to 0.1.

We used a Young's modulus of the same order of magnitude as measured from eel \citep{JohnH.Long1998} and lamprey \citep{Tytell2018} tissues, $E = 10^{5}$ Nm$^{-2}$, and a Poisson's ratio similar to that of human tissue, $\nu = 0.3$ \citep{choi2005estimation}.

\subsection{Marker data pre- \& post-processing and validation}

\subsubsection{Marker orientation offset reduction}

In experiments, even when care was taken during marker placement, markers can never be perfectly parallel to the local coronal plane of the animal body with the front edge perfectly perpendicular to the midline. Such an offset can affect interpolation results.

To reduce this offset, for each marker, we observed all the videos with the same marker placement and manually selected a reference video frame in which the snake body was visually straight around that marker and horizontal. To correct for yaw offset, we manually rotated the reference marker frame about its $x$ axis until its forward direction was visually tangent to the local snake body in this video frame. To correct for roll and pitch offsets, we calculated a correction rotation matrix $ R_{correction}  = R_{ref}^{-1}R_{yaw}$, where $R_{ref}$ is the rotation matrix of the marker in the reference video frame after yaw offset was corrected, $R_{yaw}$ is a rotation matrix with zero pitch and roll angles and the same yaw angle as the marker in the reference video frame after yaw offset was corrected. Then we rotated the marker frame in each video frame by applying the correction rotation matrix to its orientation. In this process, all Euler angles were calculated after rotating body frames to match the Euler angle convention: +$x$ axis pointing forward tangent to segment backbone curve, and +$z$ axis pointing upward along the axis of left-right symmetry of cross section.

\subsubsection{Smoothing orientation}

After reducing orientation offset, we smoothed the orientation using a window average filter covering 9 adjacent frames (0.09 seconds), by applying a mean filter MATLAB function \textit{smooth2a} \citep{Reeves2009} temporally to each element of the rotation matrices. Then we used a Singular-Value-Decomposition (SVD)-based projection method \citep{belta2002svd} to ensure that the matrix was still in the rotation matrix space.

\subsubsection{Position translation}

Finally, for each snake and each marker, we translated its body frame obtained from the marker on the dorsal surface of the snake body to the body midline, along the direction normal to the marker plane by the sum of the marker thickness and the body radius $r(s)$ at the corresponding body length to obtain end constraints of the backbone curve.

\subsubsection{Post-processing}

After the snake body backbone curve was reconstructed, the kinematic data were post-processed with a temporal linear interpolation to fill missing body frames resulting from missing tracking data, followed by a temporal-spatial window average smoothing of position data to reduce discontinuities produced by the piecewise numerical convergence process.

\subsubsection{Validation}

Finally, we projected the interpolated backbone curves of all body segments onto the experimental videos and visually examined the match to validate the reconstruction.

\begin{figure*}[!t]
  \includegraphics[width=\textwidth]{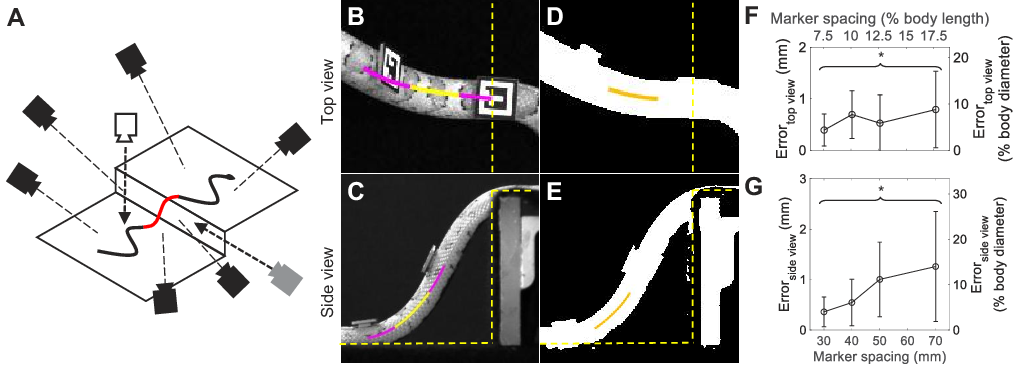}
  \caption{\textbf{Midline extraction and comparison.} (A) Experimental setup to obtain body midline ground truth. See \textbf{Experimental protocol} for description. (B, C) Representative top and side views of a snake traversing a step obstacle (yellow dashed lines) with interpolated backbone curve projected (magenta and yellow). (D, E) Representative binary top and side views with body midline projected (orange), which is extracted using computer vision techniques. (F, G) Average pointwise error of reconstructed backbone curve position from extracted midline in top and side views. Brackets and asterisks represent a significant increase of error with marker spacing ($P$ < 0.0001, linear regression). $N$ = 3 individuals, $n$ = 60 trials.}
  \label{fig_midline}
\end{figure*}

\subsection{Experiment to obtain body midline ground truth}

To characterize how marker separation affects positional accuracy of the interpolation, we performed experiments to extract the midline of a segment of the snake body between two markers (position only, Fig. \ref{fig_midline}).

\subsubsection{Experimental protocol}

We constructed a 51 mm (15\% snout-vent length) high step using extruded T-slotted aluminum and acrylic sheets (McMaster-Carr, Elmhurst, IL, USA). The entire test surfaces were covered by black felt (OnlineFabricStore, New England, USA) and lit by work lights (Designers Edge, Union City, NJ, USA) to provide high contrast of the snake body against the background. This allowed us to use computer vision techniques to extract the body midline as the ground truth.

Eight high-speed cameras (N-5A100-Gm/CXP-6-1.0, Adimec, Eindhoven, The Netherlands) recorded the test area at 100 frame/s from different views to cover the entire range of 3-D body rotation (Fig. \ref{fig_midline}A). Their videos were used to track and reconstruct 3-D position and orientation of the BEEtag markers \citep{Crall2015} on the snake body using direct linear transformation (DLT) \citep{Hedrick2008}. One camera was placed from a top view with its lens axis perpendicular to the horizontal plane (Fig. \ref{fig_midline}A, white camera). Another camera was placed from a side view with its lens axis perpendicular to the vertical plane (Fig. \ref{fig_midline}A, gray camera). Videos from these two cameras were used to extract the midline of the snake body projected into these two planes.

We used three juvenile variable kingsnakes (body length = 39.6 $\pm$ 0.4 cm, snout-vent length = 34.6 $\pm$ 0.4 cm) for midline extraction experiments. For each of the three animals, we compared our reconstructed backbone curve with the extracted midline for four different marker spacing (3, 4, 5, and 7 cm), each with 5 trials. This resulted in a total of 60 trials. All animal experiments were approved by and in compliance with The Johns Hopkins University Animal Care and Use Committee (protocol RE16A223).

\subsubsection{Midline extraction using computer vision techniques}

For both the top and side view videos, we used layered image morphology operations to extract the outline of the snake body between markers. The image operations include edging, dilation, erosion, and filling.

First, we converted each video frame (Fig. \ref{fig_midline}B, C) to a binary image with white edges and black background (Fig. \ref{fig_midline}D, E) by looking for local maxima of the gradient of gray intensity using the edge function in MATLAB. Then, we connected isolated small regions with white edges resulting from snake skin texture by enlarging their boundaries (dilation), and eroded away the boundaries of the enlarged white regions to recover their original size. Next, we filled small black regions remaining inside the snake body by making them white (filling). By only keeping the largest remaining white area between the markers, we further ensured that the two outlines (left and right for top view, dorsal and ventral for side view) extracted were clean without undesired dots.

Once the two outlines of the body were found, the projection of the midline into the horizontal and vertical planes could be recovered. We evenly interpolated the same number of points along both outlines. Then for each pair of these points, we assigned a line segment whose length is body radius normal to the corresponding outline, starting at each point and pointing into the body, and connected their endpoints with a line segment. The midpoint of this line segment was considered as the midline point corresponding to this pair of points. The extracted midline was then described by the sequence of these midline points (Fig. \ref{fig_midline} D-E, orange curve). The extracted midline was shorter than marker separation because the markers extrude beyond the body outline and were not included in midline extraction.

\subsubsection{Midline comparison}

Only the position data from our reconstructed backbone curve were used to compare with the extracted midline, which lacked orientation information. For every video frame of each trial, the reconstructed backbone curve was first projected onto the video frame (Fig. \ref{fig_midline}B, C, magenta and yellow curve) using the camera model coefficients extracted by the DLT calibration. To compare with the extracted midline that was shorter, we truncated the reconstructed backbone curve using two normal vectors intersecting both ends of the extracted midline. The truncated backbone curve  (Fig. \ref{fig_midline}B, C, yellow curve) was then linearly interpolated to have the same number of points as the extracted midline for calculation of pointwise errors. These errors were then averaged for each video frame and then across video frames weighted by the extracted midline length for each trial.

\subsection{Comparison with B-spline interpolation}

To evaluate the accuracy of our method, we compared its interpolation results with those obtained using the widely-used B-spline method. Most previous animal locomotion studies \citep{Sharpe2015,Schiebel2018,Fontaine2008} only applied B-spline interpolation to position data. Here, we applied it to position data and then calculated orientation from the interpolated position curve (see \textbf{Implementation of B-spline} and \cite{Yeaton2020}).

For this comparison, we used both our method and the B-spline method (see \textbf{Definition of B-spline}) to reconstruct the continuous body of the three kingsnakes during traversal of a large step obstacle, in which the body deforms substantially in three dimensions (both position and orientation) \citep{Gart2019}. A total of 122 trials were used for statistical analysis, with a video frame rate of 100. In each trial, 10-14 BEEtag markers \citep{Crall2015} were placed along the snake body to track local body position and orientation in 3-D. For both methods, all data were pre- and post-processed as described above.

We used each method to reconstruct a body section in the snake body, which has two markers at its two ends and another in the middle as the approximate ground truth. Then, in each interpolated curve, we selected the body frame that corresponds to the middle marker selected. We compared the interpolated with the tracked middle marker frame to obtain interpolation error for each method.

For our method, we used the front and rear of these three markers (44--73 mm apart) as end constraints to perform backbone interpolation. The middle marker in between was not used as an end constraint but to provide an approximate ground truth of local position and orientation, relative to which interpolation error was measured. We measured the length of the body segments between the front and middle markers $L_1$ and between the middle and rear markers $L_2$. We then used the interpolated body frame that lay at the same location lengthwise in the parameter space $s_{mid} = L_1$ along the interpolated backbone curve $g(s), s\in\left[0,L_1+L_2\right]$ to provide the interpolated local position and orientation $g(s_{mid})$.

For the B-spline method, we interpolated a description of the whole body $g(\sigma),\sigma \in \left[0,1\right]$ using all the markers placed along the body, except the middle one used to provide the approximate ground truth mentioned above, resulting in a total number of 9--13 markers used (variable between trials). All but the middle markers were used because B-spline interpolation will produce a straight line if only two markers are used. The middle marker mentioned above lay at a location of $\sigma_{mid}=\sigma_{k_{front}}+(\sigma_{k_{rear}}-\sigma_{k_{front}})\frac{L_1}{L_1+L_2}$ in the $\sigma$ space, where $\sigma_{k_{front}}$ and $\sigma_{k_{rear}}$ correspond to the front and rear markers (see definition of $\sigma$ in section below). We then used the interpolated body frame that lay at the same location lengthwise in the $\sigma$ space to provide the interpolated local position and orientation $g(\sigma_{mid})$.

For both methods, deviation of the interpolated local position and orientation ($g(s_{mid})$ or $g(\sigma_{mid})$) from the approximate ground truth obtained from the middle marker gave the interpolation error. We used the Euclidean distance to measure position error and the Frank Park distance \citep{Park1995} to measure orientation error.

For each video frame of each trial, each marker on the body except the first and the last one was used as the approximate ground truth once to obtain the errors of each method. Errors for all reference markers from all video frames and all trials were then pooled and averaged to obtain average error for each method.

\subsubsection{Definition of B-spline}

The B-spline method \citep{deBoor1978} uses a linear combination of B-spline basis functions to interpolate between given points, defined as follows. We use $\sigma \in [0,1]$ to parameterize the B-spline curve ${\v f}(\sigma)$, which is constructed as a weighted sum of $m+1$ pre-defined B-spline basis functions:
\begin{eqnarray} \label{eqn_bspline}
  {\v f}(\sigma) = \sum_{l=0}^{m} {\v \beta}_{l}b_{l}(\sigma)
 \end{eqnarray}
where ${\v \beta}_{l}$ is the weight vector of the $l$th basis function $b_{l}(\sigma)$, each point on the B-spline curve ${\v f}(\sigma)$ and each ${\v \beta}_{l}$ is a vector of dimension $d$, and each basis function $b_{l}(\sigma), l=0,\ldots,m$ is a scalar. These basis functions are constructed by recursion \citep{deBoor1978}.

Given a sequence of $ n+1 $ real-valued sample points ${\v x}_k, k = 0, \ldots, n $ (which were either position or orientation measurements from the tracked markers in this study), the B-spline curve is constrained to meet each sample point ${\v x}_k$ at a corresponding $\sigma_k$:
\begin{equation} \label{bspline_constraint}
 {\v f}(\sigma_k) = {\v x}_k, k = 0, \ldots, n
\end{equation}

These constraints can also be written in a matrix form:
\begin{equation} \label{bspline_constraint}
  X = B{\v \beta}
\end{equation}
where ${\v \beta} = \left[{\v \beta}_0, \ldots , {\v \beta}_m \right]^T$ is the weight matrix to compute, $X = \left[{\v x}_0, \ldots ,{\v x}_n \right]^T$ is the matrix of sample points, and $B$ is a matrix of the values of basis functions at the corresponding sample points $B_{kl} = b_{l}(\sigma_{k}), k = 0, \ldots, n, \ l = 0, \ldots, m$.

\subsubsection{Implementation of B-spline}

We first used B-spline (Eqn. (\ref{bspline_constraint})) to interpolate the position of the end constraints. We used the measured $x, y, z$ values at all the selected end constraints as sample points ${\v x}_k$ and the respective body lengths from the end constraint closest to the head to each of these end constraints as $\sigma_k$. 

Next, we calculated orientation iteratively using the interpolated position curve \citep{Bloomenthal1990}. We first calculated the orientation of the body frame closest to the head by assuming it matched the tangent of the interpolated position curve with zero roll. We then iteratively updated the orientation of each following body frame by rotating the previous body frame to match the local tangent around an axis normal to both the previous and the local tangent. We note that rotation can be calculated independent of position by directly interpolating the measured orientation \citep{belta2002svd, Pletinckx1989, Park1997}, which was widely used in temporal interpolation. However, in this case of spatial interpolation, it does not match the tangent of the interpolated position curve.

The implementation used a spline interpolation function \textit{spaps} in MATLAB Curve Fitting Toolbox with two varying parameters, tolerance $tol$ and order of basis function $p = 2m-1$. This function returns a B-spline curve ${\v f}$ that minimizes the cost function:
\begin{equation}
  \rho E({\v f}) + F(D^m{\v f})
\end{equation}
where the error function $E({\v f}) = \sum\limits_{k=0}^{n}w_k\|{\v x}_k-{\v f}(\sigma_k)\|^2$, the smoothness function $F(D^m{\v f})=\int_{min(\sigma)}^{max(\sigma)}\|D^m{\v f}(\sigma)\|^2d\sigma$, and $D^m{\v f}$ is the $m$th derivative of ${\v f}$. The smoothing parameter $\rho$ was chosen so that $E({\v f}) = tol$ \citep{Reinsch1967}. The resulting B-spline was an interpolation curve through all sample points when $tol = 0$ and was as a smoothing curve when $tol > 0$. The weight vector ${\v w} = [w_0,\ldots,w_n]^T$ was calculated from $\sigma_k$ to approximate $\int_{min(\sigma)}^{max(\sigma)}\| {\v f} - {\v x}\|^2 d\sigma$ with discrete values ${\v f}(\sigma_k), k=0,\ldots,n$:
\begin{align}
  \begin{split}
    {\v w} = [&\frac{d\sigma_0}{2}, \frac{d\sigma_0+d\sigma_1}{2}, \frac{d\sigma_1+d\sigma_2}{2}, \ldots,\\
    &\frac{d\sigma_{n-2}+d\sigma_{n-1}}{2},\frac{d\sigma_{n-1}}{2}], \\
    d\sigma_i = &\sigma_{i+1} - \sigma_i, i = 0,\ldots,n-1
  \end{split}
\end{align}

Because cubic B-spline ($m = 2$) is widely used in animal locomotion literature \citep{Sharpe2015,Fontaine2008,Yeaton2020}, we used $m = 2$ in our B-spline implementation to compare with the backbone interpolation method. We used $tol=0$ (interpolation) because this produced smaller errors than for $tol>0$ (smoothing). We performed additional comparisons by varying $m\in\{1,2,3\}$ and $tol\in\{0, 0.2, 0.4\} \times$ body length for interpolating ${\v p}$ (Fig. S3).

\subsection{Statistics}

To compare our interpolation results with the extracted midline position ground truth, for each marker spacing, weighted errors from all video frames of all trials were pooled to calculate their means and standard deviations.

To compare our method with the B-spline method, for each method, interpolation errors for all reference markers from all video frames of all trials were pooled to calculate their means and standard deviations.

To test whether the position error in top and side view planes depended on marker separation, we used a linear regression for each of these measurements, with marker separation as a continuous independent factor and each position error as a continuous dependent factor.

To test whether our method and the B-spline method differed in interpolation accuracy, we used an ANOVA for both position and orientation errors, with the method as a nominal independent factor and each error as a continuous dependent factor.

All the statistical tests followed the SAS examples in \citep{McDonald2014} and were performed using JMP Pro 13 (SAS Institute, Cary, NC, USA).

\section{Results}

\begin{figure}[!h]
  \centering
  \includegraphics[width=3.5in]{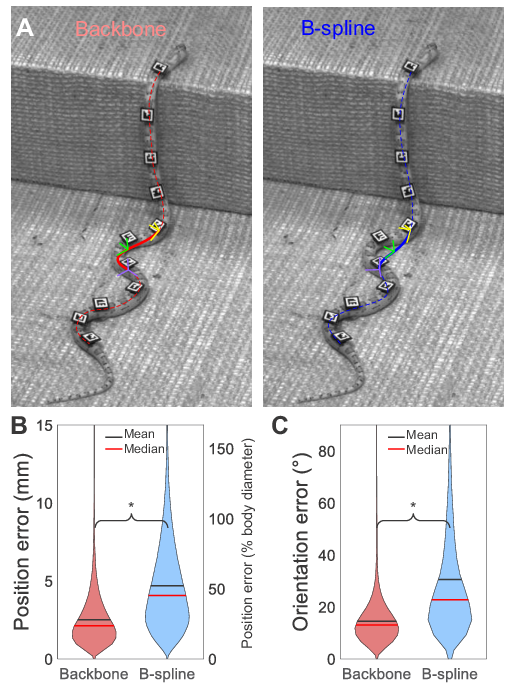}
  \caption{\textbf{Comparison of interpolation accuracy between our method and B-spline method.} (A) A representative snapshot of interpolation results. See Movie 3 for a representative trial. (B) Position error. (C) Orientation error. Data are shown using violin plots. Black and red lines show mean and median. Local width of graph is proportional to frequency of data along $y$ axis. Brackets and asterisks represent a significant difference ($P$ < 0.0001, ANOVA). $N$ = 3 individuals, $n$ = 122 trials.}
  \label{fig_result}
\end{figure}

Comparison of the interpolated backbone curve with the extracted midline via computer vision techniques showed that our interpolation method achieved high positional accuracy. The average position error over the snake body segment reconstructed (3-7 cm) in both top and side views were within 1.5 mm or 17\% of body diameter (9 mm) (Fig. \ref{fig_midline}F-G). In addition, as marker separation became smaller, position error decreased in both top ($P < 0.0001$, $R^2 = 0.03$, linear regression) and side ($P < 0.0001$, $R^2 = 0.17$, linear regression) view planes. Thus, users should use the smallest marker separation possible that does not significantly affect the animal's behavior.

Our interpolation method provided a good approximation of the kingsnake's continuous body with both position and orientation information while it uses large 3-D body deformation to traverse a large step obstacle which, when combined with body surface reconstruction, enabled body-terrain contact estimation (Movie 2). Compared with the widely-used cubic B-spline interpolation method (with cubic B-spline basis functions and zero tolerance, Fig. \ref{fig_result}A, blue), our method (Fig. \ref{fig_result}A, red) achieved higher interpolation accuracy in both position and orientation, by nearly a 50\% reduction in error (Fig. \ref{fig_result}B, C, $P < 0.0001$, ANOVA; Movie 3). In addition, varying the degree of basis functions and tolerance for the B-spline method did not reduce its error to as small as that of our method (Fig. S3B).

This high accuracy demonstrated that, when constrained by closely spaced end constraints, an over-simplified elastic rod can still achieve a higher accuracy than purely geometric methods can for interpolating the continuous body of an limbless animal body with complex morphology and biomechanics.

Note that the average errors of the two methods are larger than that of our method from the extracted midline. This may be due to three reasons. First, when we compared the two methods, the tracked marker used as reference can introduce measurement error itself. Second, error is usually larger in the middle of each interpolated segment than near its ends. Third, in this dataset, besides a small, high friction step similar to the one used in the midline extraction experiment, the snake also traversed more challenging, larger steps and low friction steps. In these conditions, the body bent more substantially \citep{Gart2019}, which may lead to greater interpolation error.

\section{Discussion}

The higher accuracy of our interpolation method than the purely geometric B-spline method is attributed to its physical representation of the long, slender body as an elastic rod. For example, by using stiffness values that are in the ballpark from similar limbless animals, our method does not generate unrealistically large extension or compression, which the purely geometric B-spline method may produce (Fig. \ref{fig_result}A). Such a higher accuracy of our method is particularly useful for quantifying body-terrain contact (e.g., calculating the ground base of support (Movie 2 of \cite{Gart2019})), especially as position error reduced from about half to a quarter body diameter.

In addition, the backbone curve description using finite elements provides more flexibility in representing a complex shape than using a limited set of basis functions. For example, we can use different stiffnesses and cross-sectional shapes and consider tapering for interpolating different body segments or even for different finite elements within a body segment.

The higher accuracy of our method came at a cost. First, markers that can provide 3-D orientation information are usually rigid and larger than non-invasive point markers, so they take more effort to attach than painting point markers and may be difficult or impossible to use on small animals. Multiple aggregated point markers or skin texture may serve as an alternative to provide local 3-D orientation information, although this adds extra effort. In addition, the computation speed of our method was two orders of magnitude lower than that of the B-spline method (2 vs. 0.02 sec on average to reconstruct a segment with 700 finite elements in one video frame), mainly because of the inverse kinematics iterations used. Further, it takes the experimenter more effort to measure and fine-tune segment lengths, which is not necessary for the B-spline method. Finally, there are also small discontinuities between adjacent body segments produced by the piecewise inverse kinematics iterations with a finite threshold of convergence, although this can be resolved by smoothing in post-processing. Thus, users should weigh the benefits against these costs before applying our method. If 3-D position and orientation information are desired, but a lower accuracy is acceptable to save time, users may first try the B-spline method with orientation added following our implementation.

When applied to other species, several things should be considered. First, different species can have large differences in body geometry and stiffness properties. Users should measure length, cross-sectional shape, and tapering and choose marker spacing based on body dimensions and aspect ratio. For stiffness, they may need to tune Young's modulus and Poisson's ratio, as well as use a non-circular shape to calculate second moments of inertia and cross-sectional area. In addition, maximal inverse kinematics iteration steps may be adjusted to achieve a balance between convergence success rate and reconstruction time. Furthermore, thresholds used for the writhe and twist check depend on how much the animal can deform its body.

In addition to locomotion research, our method is useful for other studies that require quantification of continuous shape of limbless biological and artificial systems in three dimensions, such as snake predation using constriction \citep{Penning2017}, snake thrashing \citep{Danforth2020}, root nutation \citep{Ozkan-Aydin2019}, and octopus arm \citep{Zelman2013} and continuum robot \citep{Laschi2012} manipulation.


\begin{acknowledgements}
  We thank Henry Astley, Bob Full, and Rajat Mittal for discussion; Sean Gart for providing snake step traversal experimental data for accuracy comparison with B-spline; and Henry Astley for advice on animal care and experiments.
\end{acknowledgements}

\appendix
\section{Data availability}
We uploaded MATLAB codes and demonstration files to a GitHub repository: \url{https://github.com/TerradynamicsLab/continuous_body_3D_reconstruction.git}

\section{Supplementary materials}
Three supplementary figures are uploaed as ancillary files. Three supplementary movies are uploaded to YouTube:

\href{https://youtu.be/QsGgHHXlHrs}{\textbf{Movie 1.}} Stick figure is insufficient to quantify body-terrain interaction.

\href{https://youtu.be/vD5x_NuuxEo}{\textbf{Movie 2.}} Continuous body 3-D reconstruction using backbone interpolation.

\href{https://youtu.be/u6sI9J9nqpU}{\textbf{Movie 3.}} Comparison between backbone and B-spline interpolation. In this example, the sixth marker from the head is used as an approximate ground truth and not used for interpolation. Comparison of interpolation error from this ground truth between backbone and B-spline methods is shown in Fig. 3.




\bibliographystyle{unsrt}
\bibliography{snake_modeling.bib}

\end{document}